\documentclass[12pt,draftcls,onecolumn]{IEEEtran}
\usepackage{setspace}
\usepackage{amsmath,amsfonts}
\usepackage{algorithmic}
\usepackage{algorithm}
\usepackage{array}
\usepackage{color}
\usepackage{xcolor}
\usepackage[caption=false,font=footnotesize,labelfont=rm,textfont=rm]{subfig}
\usepackage{textcomp}
\usepackage{stfloats}
\usepackage{url}
\usepackage{verbatim}
\usepackage{graphicx}
\usepackage{cite}
\usepackage{amssymb}
\usepackage{amsmath,amsfonts}
\usepackage{url}
\begin{document}

\title{\huge Stacked Intelligent Metasurface Enabled LEO Satellite Communications Relying on Statistical CSI}

\author{Shining Lin, Jiancheng An,~\IEEEmembership{Member,~IEEE,} Lu Gan,\\M\'erouane Debbah,~\IEEEmembership{Fellow,~IEEE,} and Chau Yuen,~\IEEEmembership{Fellow,~IEEE}
\vspace{-0.4in}
 % <-this % stops a space
\thanks{This work is partially supported by Sichuan Science and Technology Program under Grant 2023YFSY0008 and 2023YFG0291, and partially supported by Yibin Science and Technology Program under Grant YBP-002. The research of Chau Yuen is supported by the Ministry of Education, Singapore, under its MOE Tier 2 (Award number MOE-T2EP50220-0019).}
\thanks{S. Lin and L. Gan are with the School of Information and Communication Engineering, University of Electronic Science and Technology of China, Chengdu, Sichuan, 611731, China. L. Gan is also with the Yibin Institute of UESTC, Yibin 644000, China (e-mail: 202221011710@std.uestc.edu.cn; ganlu@uestc.edu.cn).}% <-this % stops a space
\thanks{J. An and C. Yuen is with the School of Electrical and Electronics Engineering, Nanyang Technological University, Singapore 639798 (e-mail: jiancheng\_an@163.com; chau.yuen@ntu.edu.sg).}
\thanks{M. Debbah is with KU 6G Research Center, Khalifa University of Science and Technology, Abu Dhabi 127788, UAE (e-mail: merouane.debbah@ku.ac.ae).}}

% The paper headers
\markboth{DRAFT}%
{Shell \MakeLowercase{\textit{et al.}}: A Sample Article Using IEEEtran.cls for IEEE Journals}

% Remember, if you use this you must call \IEEEpubidadjcol in the second
% column for its text to clear the IEEEpubid mark.

\maketitle

\begin{abstract}
Low earth orbit (LEO) satellite communication systems have gained increasing attention as a crucial supplement to terrestrial wireless networks due to their extensive coverage area. This letter presents a novel system design for LEO satellite systems by leveraging stacked intelligent metasurface (SIM) technology. Specifically, the lightweight and energy-efficient SIM is mounted on a satellite to achieve multiuser beamforming directly in the electromagnetic wave domain, which substantially reduces the processing delay and computational load of the satellite compared to the traditional digital beamforming scheme. To overcome the challenges of obtaining instantaneous channel state information (CSI) at the transmitter and maximize the system's performance, a joint power allocation and SIM phase shift optimization problem for maximizing the ergodic sum rate is formulated based on statistical CSI, and an alternating optimization (AO) algorithm is customized to solve it efficiently. Additionally, a user grouping method based on channel correlation and an antenna selection algorithm are proposed to further improve the system performance. Simulation results demonstrate the effectiveness of the proposed SIM-based LEO satellite system design and statistical CSI-based AO algorithm. 
\end{abstract}

\section{Introduction}
\IEEEPARstart{R}{ecently}, the International Telecommunication Union-Radiocommunication Sector (ITU-R) has identified global coverage as a crucial objective for designing sixth-generation (6G) networks. As one of the important technologies to achieve this goal, low earth orbit (LEO) satellite communication is gaining increasing attention due to its wide coverage area, relatively low costs, reduced latency, and power consumption compared to geostationary earth orbit (GEO) satellites. In industry, companies such as OneWeb and SpaceX have launched projects to deploy large-scale constellations of multiple LEO satellites\cite{arXiv_2023_An_Toward, kodheli2020satellite}.

To achieve higher communication capacity, the multi-beam satellite technique, which generates multiple pencil beams to serve multiple users simultaneously, has been widely adopted\cite{kodheli2020satellite}. The multi-beam technique is generally implemented by using digital precoding to effectively reduce inter-beam interference \cite{schwarz2019mimo,li2021downlink}. However, digital precoding techniques require installing a high-speed baseband digital signal processor at the satellite, which increases the satellite's power consumption and computational demand.

Additionally, most of the existing precoding schemes in satellite communication systems assume the availability of accurate instantaneous channel state information (CSI). However, in LEO satellite systems with high-speed mobility and long propagation delay, obtaining accurate instantaneous CSI is challenging. To address this, researchers have developed robust precoding methods based on statistical CSI\cite{li2021downlink,TGCN_2022_An_Joint, chen2022robust}, which varies relatively slowly and is easier to obtain, to maximize the ergodic sum rate.

Parallelly, reconfigurable intelligent surface (RIS) has shown its great potential to enhance spectrum and energy efficiencies. Specifically, a RIS is an artificial two-dimensional (2D) metasurface that is capable of controlling the electromagnetic (EM) behavior of incident waves using low-cost meta-atoms\cite{AnChannel}. By appropriately optimizing the reflection pattern, RIS has achieved substantial capacity improvements even with moderate signaling overhead\cite{AnCodebook} and reduced computational complexity\cite{AnAntenna}. However, current RIS designs generally adopt a single metasurface layer, which constrains the tuning capability for practical meta-atoms. To address this issue, an advanced technology based on stacked intelligent metasurface (SIM) has emerged recently. In\cite{an2023stacked}, \emph{An et al.} integrated SIM with the transmit antenna array, enabling beamforming entirely in the wave domain, which significantly reduces the processing delay and energy consumption compared to its digital counterpart. Subsequently, they extended this integration to the receive antenna array, enabling SIM to implement both transmit precoding and receiver combining functions\cite{AnSIM}. However, the research on SIM-aided communication systems still has numerous gaps\cite{an2023stackeds}. For instance, further evaluating the energy efficiency of SIM-aided systems may constitute a promising topic, which may require the accurate modeling of the energy loss of EM wave propagating through the SIM.

Motivated by this advanced SIM technology and the requirements of traditional digital precoding techniques on satellite onboard capabilities, in this letter, we propose an SIM-based multiuser multiple input single output (MISO) satellite communication system. In this system, the signal for each user terminal (UT) is transmitted directly from a corresponding transmit antenna, and beamforming is entirely accomplished in the wave domain by the SIM integrated with the transmitter. With this approach, the digital precoding is completely removed from the satellite, substantially reducing the processing delay and energy consumption incurred in precoding operations on high-rate baseband data during the downlink transmission from the satellite to UTs. Additionally, to reduce the pilot overhead for probing channels, slow-varying statistical CSI is utilized. Specifically, our major contributions are summarized as follows: First, based on the statistical CSI, we formulate an optimization problem to maximize the ergodic sum rate by designing the phase shifts in the SIM and power allocation at the transmitter, and customize an efficient alternating optimization (AO) algorithm to solve it. Furthermore, we propose a user grouping method for reducing the inter-channel correlation and design an antenna selection algorithm based on the minimization of the total leakage energy to improve the performance of LEO satellite communications when serving a large number of users in its coverage. The simulation results demonstrate that even without digital precoding, SIM-based systems can achieve performance comparable to digital systems. They also show that using statistical CSI approaches the performance of instantaneous CSI, and the gains over existing methods are attributed to the devised grouping and antenna selection methods.
\section{System Model}
Consider a LEO satellite communication system where an SIM-based multi-antenna satellite serves $K$ single-antenna UTs simultaneously. As shown in Fig. 1, the satellite is equipped with a uniform planar array consisting of $M\ge K$ antennas and with an SIM used for carrying out beamforming in the wave domain. The SIM is composed of $L$ parallel metasurface layers, each consisting of $N$ meta-atoms. Remarkably, SIM achieves transmit beamforming naturally by imposing the phase shifts on the EM waves as they pass through each meta-atom, under the control of a smart controller attached to the SIM. Unlike traditional precoding schemes, in an SIM-based system, each antenna transmits the data stream intended for a single UT. This requires selecting $K$ from $M$ antennas, which will be detailed in Section IV-B.

\begin{figure}[!t]
\centering
\includegraphics[width=2.6in]{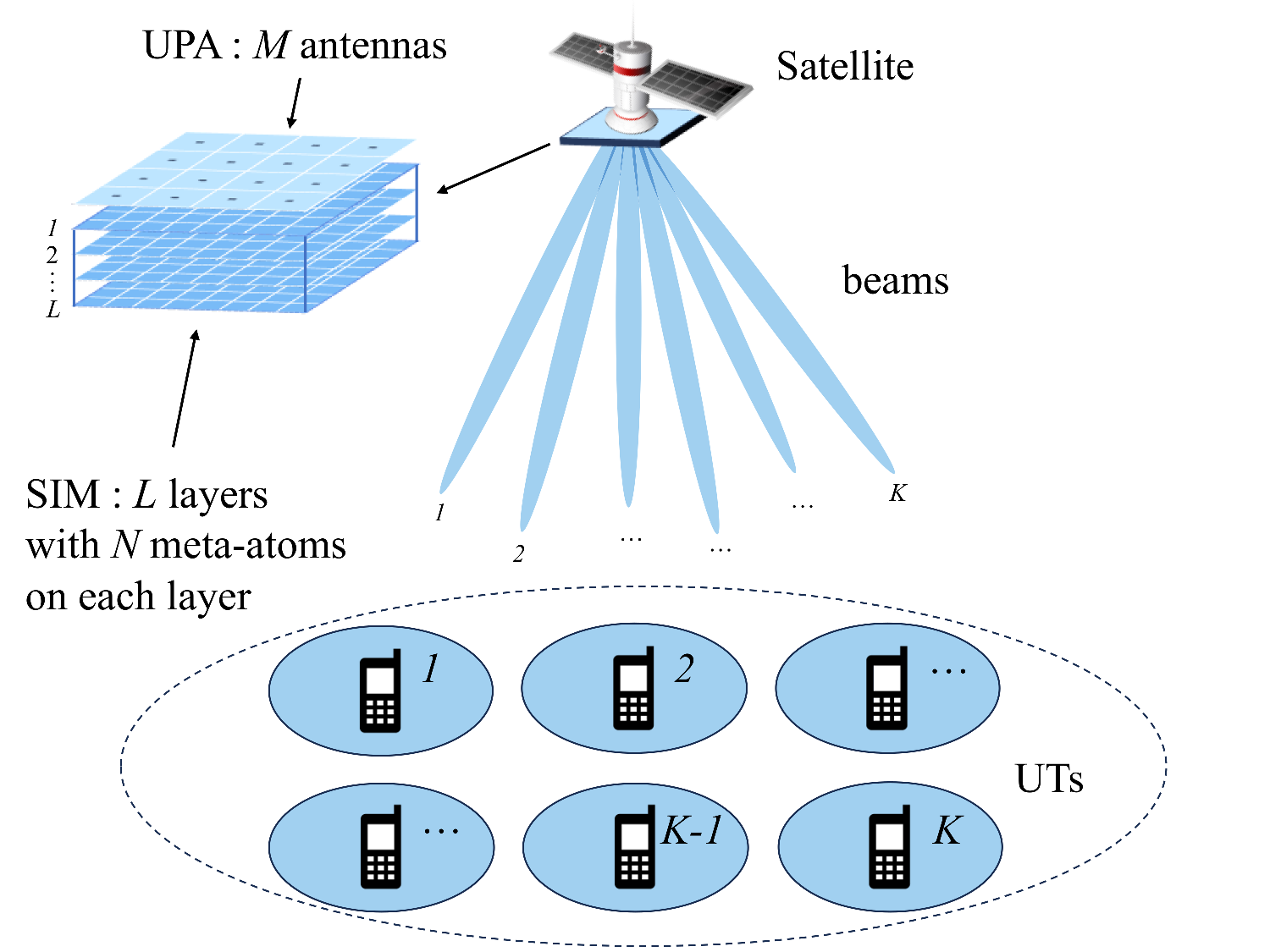}
\caption{An SIM-based MISO satellite network.}
\label{fig_1}
\end{figure}

Let $\mathcal{L}=\{1,2,\cdots,L\}$, $\mathcal{N}=\{1,2,\cdots,N\}$, $\mathcal{K}=\{1,2,\cdots,K\}$ denote the set of metasurfaces, meta-atoms on each metasurface layer, and UTs, respectively. Moreover, let ${\bf\Theta} \in{\mathbb R}^{N \times L}$ denote the phase shifts in the SIM and its $(n,l)$-th entry $\theta^l_n \in (0,2\pi], \forall l\in\mathcal{L}, \forall n\in\mathcal{N}$ denotes the phase shift imposed by the $n$-th meta-atom on the $l$-th metasurface layer. Furthermore, let ${\bf\Phi}^l = {\text {diag}}(e^{j\theta^l_1},e^{j\theta^l_2},\cdots,e^{j\theta^l_N})\in{\mathbb C}^{N \times N}$ denote the phase shift matrix induced by the $l$-th metasurface layer. Additionally, let ${\bf W}^l\in{\mathbb C}^{N \times N},\forall l\ne1,l\in\mathcal{L}$ denote the propagation matrix from the $(l-1)$-th metasurface layer to the $l$-th metasurface layer. The $(n,n^{\prime})$-th entry of ${\bf W}^l$ represents the attenuation coefficient from the $n^{\prime}$-th meta-atom on the $(l-1)$-th metasurface layer to the $n$-th meta-atom on the $l$-th metasurface layer. According to the Rayleigh-Sommerfeld diffraction theory\cite{lin2018all, arXiv_2024_An_Two}, $w^l_{n,n^{\prime}}$ is given by $w^l_{n,n^{\prime}} = {\frac{d_xd_y{\text{cos}}{\alpha}^l_{n,n^{\prime}}}{d^l_{n,n^{\prime}}}}\left({\frac{1}{2\pi d^l_{n,n^{\prime}}}}-j{\frac{1}{\lambda}}\right)e^{j2\pi{d^l_{n,n^{\prime}}}/\lambda}$, where $d_x \times d_y$ is the size of each meta-atom, ${\alpha}^l_{n,n^{\prime}}$ denotes the angle between the propagation direction and the normal direction of the $(l-1)$-th metasurface layer, $d^l_{n,n^{\prime}}$ denotes the propagation distance, and $\lambda$ is the radio wavelength. Additionally, let ${\bf w}_k\in{\mathbb C}^{N \times 1}$ denote the attenuation coefficient from the antenna required to transmit the signal for the $k$-th UT to the first metasurface layer, and its $n$-th entry $w_{n,k}$ can also be obtained using the same calculation method as $w^l_{n,n^{\prime}}$. Therefore, the beamforming matrix of SIM ${\bf G}\in{\mathbb C}^{N \times N}$ can be written as ${\bf G}={\bf\Phi}^L{\bf W}^L{\bf\Phi}^{L-1}{\bf W}^{L-1}\cdots{\bf\Phi}^2{\bf W}^2{\bf\Phi}^1$.

In this letter, we employ the Rician channel model to characterize the LEO satellite communications. Additionally, the path loss is modeled by assuming free space propagation. Therefore, the channel vector from the last layer of SIM to the $k$-th UT is written as\\
${\bf h}_k^H=\sqrt{\beta_k}\left(\sqrt{\frac{\kappa_k}{1+\kappa_k}}{\bf h}^H_{k,{\text{LoS}}}+ \sqrt{\frac{1}{1+\kappa_k}}{\bf h}^H_{k,{\text{NLoS}}}\right)\in{\mathbb C}^{1 \times N}$, where $\kappa_k$, $\beta_k$, ${\bf h}_{k,{\text{LoS}}}\in{\mathbb C}^{N \times 1}$, and ${\bf h}_{k,{\text{NLoS}}}\in{\mathbb C}^{N \times 1}$ represent the Rician factor, channel gain, LoS component, and NLoS component of the $k$-th channel, respectively. Moreover, $\beta_k=g^{\text{tra}}g_k^{\text{rec}}{(\frac{\lambda}{4\pi r_k})}^2$, where $g^{\text{tra}}$ is the transmit antenna gain, $g_k^{\text{rec}}$ is the $k$-th UT receive antenna gain and $r_k$ is the distance between SIM and the $k$-th UT, ${\bf h}_{k,{\text{LoS}}}$ represents the response vector of the last layer of SIM and ${\bf h}_{k,{\text{NLoS}}}\in\mathcal{CN}({\bf0},{\bf \Sigma}_k)$, where ${\bf \Sigma}_k$ can be approximated as ${\bf h}_{k,{\text{LoS}}}{\bf h}_{k,{\text{LoS}}}^H$ due to the fact that satellite's height is significantly greater than that of scatterers\cite{li2021downlink}. In the following, we assume instantaneous CSI $\mathbf{h}_k^{H}$ can not be obtained while the parameters $\kappa_k$, $\beta_k$, and ${\bf h}_{k,{\text{LoS}}}$ are available and presumed constant within a coherence block.

The information symbol intended for the $k$-th UT is denoted by $s_k$, and we assume that they are mutually independent. Let $p_k$ denote the power allocated to the $k$-th UT, subject to the sum power constraint $\sum_{k=1}^{K}{p_k}=P_T$, where $P_T$ is the total transmit power available at the satellite. Therefore, the received signal $y_k$ at the $k$-th UT is given by $y_k={{\bf h}_k^H}{\bf G}\sum_{k^{\prime}=1}^K{{\bf w}_{k^{\prime}}}\sqrt{p_{k^{\prime}}}s_{k^{\prime}}+n_k$, where $n_k \sim \mathcal{CN}(0,{\sigma}_k^2)$ is the i.i.d. additive white Gaussian noise at the $k$-th UT receiver. Moreover, ${\sigma}_k^2 = k_BBT_k$ is the noise power, where $k_B=1.380649\times10^{-23}J/K$ is Boltzmann constant, $B$ is the operating bandwidth, and $T_k$ is the noise temperature.
\section{DL Transmissions Based on Statistical CSI}
\subsection{Problem Formulation}
In this letter, we aim to maximize the ergodic sum rate of $K$ UTs by optimizing $\bf \Theta$ and $\bf p$ using statistical CSI, namely ${\bf h}_{k,{\text{LoS}}}$ and $\beta_k$. However, due to the complexity of obtaining the expression for the ergodic sum rate, we define the parameter $\overline{\gamma}_k=\frac{{\mathbb E}\left[|{\bf h}_k^H{\bf G}{\bf w}_k|^2p_k\right]}{{\mathbb E}\left[\sum ^K_{k^{\prime}\ne k}|{\bf h}_k^H{\bf G}{\bf w}_{k^{\prime}}|^2p_{k^{\prime}}+\sigma^2_k\right]}=\frac{\beta_k|{{\bf h}_{k,{\text{LoS}}}^H}{\bf G}{\bf w}_k|^2p_k}{\sum ^K_{k^{\prime}\ne k}\beta_k|{\bf h}^H_{k,{\text{LoS}}}{\bf G}{\bf w}_{k^{\prime}}|^2p_{k^{\prime}}+\sigma^2_k}$ as the average signal to interference plus noise ratio for the $k$-th user to simplify the problem. Specifically, the simplified problem is formulated as 
\begin{subequations}
\begin{align}
&&{\underset{\bf{\Theta},\bf {p}} {\text{max}}}&\quad{\overline{R}}=\sum_{k=1}^K\log(1+\overline{\gamma}_k)&\label{4a}\\
&&\text{s.t.}&\sum_{k=1}^{K}{p_k}=P_T,\label{4b}&\\
&&&p_k\ge0\label{4c}, \forall k\in\mathcal{K},&\\
&&&\theta^l_n \in (0,2\pi], \forall l\in\mathcal{L}, \forall n\in\mathcal{N}.\label{4d}&
\end{align}
\end{subequations}
\noindent It can be proven that ${\overline{R}}$ in (1a) is an upper bound for ergodic sum rate due to the high channel correlation in the satellite communication system\cite{li2021downlink}. However, solving the non-concave optimization problem (1) involving variable coupling between $\bf\Theta$ and $\bf p$ is still challenging. Therefore, we next customize an AO algorithm to solve it.
\subsection{The Proposed AO Algorithm}
In this subsection, we decompose problem (1) into two sub-problems: optimizing $\bf p$ to maximize $\overline{R}$ under given $\bf\Theta$, and optimizing $\bf\Theta$ to maximize $\overline{R}$ under given $\bf p$. They are solved iteratively and the values of $\bf p$ and $\bf\Theta$ are successively updated at each iteration. The outer loop progresses until convergence or reaching the maximum number of iterations.

\subsubsection{Optimization of the Power Allocation $\bf p$ under Given $\bf\Theta$} \noindent When $\bf\Theta$ is given, the original problem is reduced to a power allocation problem considering interference from other channels under a total power constraint. The iterative water-filling algorithm, with computational complexity $O(K)$ per iteration, can progressively increase the sum rate\cite{jindal2005sum}. Furthermore, we introduce a weighted summation step by superimposing the updated solution with the previous iteration's result to enhance the stability of the algorithm. Specifically, in each iteration, the power allocation vector $\bf p$ is updated iteratively as
\begin{equation}
 \begin{aligned}
\label{eq7}
p_k \gets& \frac{1}{K}\left(p_{\text{th}}-\frac{\sum ^K_{k^{\prime}\ne k}\beta_k|{\bf h}^H_{k,{\text{LoS}}}{\bf G}{\bf w}_{k^{\prime}}|^2p_{k^{\prime}}+\sigma^2_k}{\beta_k|{\bf h}^H_{k,{\text{LoS}}}{\bf G}{\bf w}_k|^2}\right)^+\\
&+ \left(1-\frac{1}{K}\right)p_k ,\forall k\in\mathcal{K},
\end{aligned}
\end{equation}
\noindent where $p_{\text{th}}$ denotes the water-filling power threshold computed in each iteration satisfying the total power constraint and $(x)^+\triangleq\text{max}\{0,x\}$.
 \subsubsection {Optimization of the SIM Phase Shifts $\bf\Theta$ under Given $\bf p$}
\noindent Due to the high degree of coupling among the optimization variables, problem (1) is still a challenging non-convex optimization problem even if $\bf p$ is given. In this letter, we adopt the gradient ascent method. Specifically, after random initialization, the phase shift $\theta^l_n$ is updated iteratively as
\begin{equation}
\label{eq10}
\theta^l_n \gets \theta^l_n+\mu\frac{\partial{\overline{R}}}{\partial{\theta^l_n}},\forall l\in\mathcal{L}, \forall n\in\mathcal{N},\\
\end{equation}
\noindent where $\mu>0$ is the Armijo step size obtained by using the backtracking line search to ensure that each iteration increases the sum rate. Due to the constraints on transmit power imposing an upper bound on the sum rate, the algorithm can be guaranteed to converge to a local maximum. The component $\frac{\partial{\overline{R}}}{\partial{\theta^l_n}}$ of gradient is given by

\begin{equation}
\label{eq9}
\begin{aligned}
\frac{\partial{\overline{R}}}{\partial{\theta^l_n}} =& 2\log_2e\sum_{k=1}^K\left[\frac{\sum_{k^{\prime}=1}^K\eta_{k,k^{\prime}}p_{k^{\prime}}}{\sum_{k^{\prime}=1}^K\alpha_{k,k^{\prime}}p_{k^{\prime}}+\sigma_k^2}\right.\\
&\left.-\frac{\sum_{k^{\prime}\ne k}^K\eta_{k,k^{\prime}}p_{k^{\prime}}}{\sum_{k^{\prime}\ne k}^K\alpha_{k,k^{\prime}}p_{k^{\prime}}+\sigma_k^2}\right],
\end{aligned}
\end{equation}
\noindent where $\alpha_{k,k^{\prime}}=\beta_k|{\bf h}^H_{k,{\text{LoS}}}{\bf G}{\bf w}_{k^{\prime}}|^2$ and $\eta_{k,k^{\prime}}=\frac{\partial{\alpha_{k,k^{\prime}}}}{\partial{\theta^l_n}}=2\beta_k\text{Im}\{e^{-j\theta^l_n}{\bf w}_{k^{\prime}}^H{{\bf a}}^l_n({\bf b}_n^l)^H{\bf h}_{k,{\text{LoS}}}{\bf h}^H_{k,{\text{LoS}}}{\bf G}{\bf w}_{k^{\prime}}\}$. Considering $N$ is typically much larger than $L$ and $K$, the computational complexity is dominated by calculating ${\bf G}$, resulting in a complexity of $O(LN^3)$ for each gradient update and step length search. Furthermore, ${{\bf a}}^l_n$ and ${\bf b}_n^l$ represent the $n$-th column of the matrix ${\bf A}_l^H\in{\mathbb C}^{N \times N}$ and ${\bf B}_l\in{\mathbb C}^{N \times N}$, respectively. The specific expressions of ${\bf A}_l$ and ${\bf B}_l$ can be given by
\begin{align}
{\bf A}_l&=\begin{cases}{\bf W}^l{\bf\Phi}^{l-1}\cdots{\bf\Phi}^{2}{\bf W}^2{\bf\Phi}^{1},\quad &\text{if}\ l\neq 1,\\ {\bf I}_N,\quad &\text{if}\ l= 1.\end{cases}\\
{\bf B}_l&=\begin{cases}{\bf\Phi}^{L}{\bf W}^L{\bf\Phi}^{L-1}\cdots{\bf\Phi}^{l+1}{\bf W}^{l+1},\quad &\text{if}\ l\neq L,\\ {\bf I}_N,\quad &\text{if}\ l= L.\end{cases}
\end{align}

\section{ Proposed User Grouping and Antenna Selection Methods}
\subsection{User Grouping}
In practice, the total number of UTs to be serviced $K_{\text{tot}}$ in an area is much larger than $M$. Therefore, we divide UTs into $G$ groups, each containing $K_g$ UTs and using SIM-based space division multiple access within the same group and time division multiple access between different groups. Furthermore, let $\mathcal{K}_{\text{tot}}=\{1,2,\cdots,K_{\text{tot}}\}$, ${\mathcal{G}}=\{{\mathcal{K}}_1,{\mathcal{K}}_2,\cdots,{\mathcal{K}}_G\}$, $\mathcal{K}_g=\{k_{g,1},k_{g,2},\cdots,k_{g,K_g}\},k_{g,i}\in\mathcal{K}_{\text{tot}}$ denote the set of total UTs, UT groups, and UTs assigned to the $g$-th group, respectively. Before further elaborating on the user grouping algorithms, we explain several key considerations. First, in order to ensure fairness, each user is assigned to be within only one group. Moreover, to maximize the average throughput, the sum rate of each group should be maximized while minimizing the number of groups since each group occupies a time slot. 

In\cite{schwarz2019mimo}, it is demonstrated that the channels in the capacity-optimal LoS-MIMO system are mutually orthogonal. Following this philosophy, in this letter, we group users that are nearly orthogonal together. The degree of orthogonality between two channel vectors is measured by the correlation of coefficient (CoC), which is defined as ${\cos{(\measuredangle({\bf h}_i,{\bf h}_j))}}=\frac{|{\bf h}_i^H{\bf h}_j|}{\left\|{\bf h}_i\right\|\left\|{\bf h}_j\right\|}\in (0,1]$. A smaller CoC value indicates stronger orthogonality between the channel vectors. The user grouping based on CoC is computationally efficient compared to the exhaust algorithms. Aiming at minimizing the maximum value of CoC within each group, the user grouping problem can be formulated as
\begin{subequations}
\begin{align}
&&{\underset{\mathcal{G}} {\text{min}}}&\ {\underset{\mathcal{K}_g} {\text{max}}}\ {\underset{i\ne j} {\text{max}}}\ \cos{(\measuredangle({\bf h}_i,{\bf h}_j))},\ i,j\in{\mathcal{K}_g}&\label{10a}\\
&&\text{s.t.}&\bigcup_{g=1}^{G}{\mathcal{K}_g}=\mathcal{K}_{\text{tot}},\label{10b}&\\
&&&{\mathcal{K}_{i^{\prime}}}\cap{\mathcal{K}_{j^{\prime}}}=\varnothing\label{10c},\ \forall {\mathcal{K}_{i^{\prime}}},{\mathcal{K}_{j^{\prime}}}\in\mathcal{G},i^{\prime}\ne j^{\prime}.&
\end{align}
\end{subequations}
\noindent In this letter, we employ a greedy algorithm to solve problem (7) as demonstrated in \cite{hu2017user}. Additionally, we use $\sqrt{\beta_k}{\bf h}_{k,{\text{LoS}}}$ instead of instantaneous CSI to estimate the CoC. 
\subsection{Antenna Selection}
In the SIM-based LEO satellite networks, each antenna transmits the signal intended for a single UT. Hence, we need to select a unique antenna for each UT within the same UT group. In this letter, we propose an antenna selection algorithm that minimizes the sum of initial energy leakage under the assumption of uniform power allocation. The algorithm follows these steps: first, we generate $I$ beamforming matrices of SIM denoted as ${\bf{G}}_i$, $i=1,2,\cdots,I$, by applying random phase shifts multiple times. Then, for each ${\bf{G}}_i$, $i=1,2,\cdots,I$, we find an antenna selection scheme that minimizes the sum of energy leakage. Specifically, let $\mathcal{M}=\{1,2,\cdots,M\}$, ${\bf P}^{i}\in{\mathbb C}^{K \times M}$ denote the set of transmit antennas and the leakage energy matrix from ${\bf{G}}_i$, respectively. The $(k,m)$-th entry $p^{i}_{k,m}$ of ${\bf P}^{i}$ represents the sum of energy transmitted by the $m$-th antenna to non-target users when sending the signal to the $k$-th user, which is defined as $p^{i}_{k,m}=\sum ^K_{k^{\prime}\ne k}\beta_{k^{\prime}}|{\bf h}^H_{k^{\prime},{\text{LoS}}}{\bf{G}}_i{\bf w}_m|^2$. Therefore, we aim to find a set $\mathcal{M}^{i}=\{m^{i}_1,m^{i}_2,\cdots,m^{i}_K\},m^{i}_k\in\mathcal{M}$, satisfying $m^{i}_{k}\ne m^{i}_{k^{\prime}} ,\forall {k}\ne k^{\prime} $ that minimizes the total leaked energy $P^{i}_{\text{leak}}=\sum ^K_{k=1}p^{i}_{k,m^i_k}$. This problem can be efficiently solved using the Hungarian algorithm \cite{kuhn1955hungarian}. Finally, select $\mathcal{M}^{i}$ corresponding to the minimum $P^{i}_{\text{leak}}$ as the ultimate antenna selection scheme. 
\section{Simulation Results}
In this section, we provide simulation results to evaluate the performance of the SIM-based LEO satellite communication system. We consider a LEO satellite at an altitude of 1000 km, serving a circular region on the ground with a diameter of 600 km and it is assumed that the probability distributions of the elevation and azimuth angles of the satellite relative to each UT are independently and uniformly distributed. Additionally, the Earth's curvature is considered when calculating the propagation distance. The distance from the transmit antenna to the last metasurface layer is set to $5\lambda$. Therefore, the vertical spacing from the antenna array to the first metasurface and that between adjacent metasurfaces are $5\lambda/L$. The side length of each meta-atom and the spacing between adjacent antennas are set to $\lambda/2$ and $\lambda$, respectively. Similar to the simulations in \cite{li2021downlink}, the carrier frequency, transmit antenna gain, UT receive antenna gain, processing bandwidth, and noise temperature are set to $4$ GHz, $g^{\text{tra}}=6$ dBi, $g_k^{\text{rec}}=0$ dBi, $\forall k\in\mathcal{K}$, $B=50$ MHz, and $T_k=290$ K, $\forall k\in\mathcal{K}$, respectively. The Rician factors for each UT rely on the specific elevation angle\cite{3GPPTR, CL_2023_An_A}. Unless otherwise specified, the remaining parameters are set as follows: $P_T=30$ dBW, $N=225$, $L=4$, $M=K=9$, and it is assumed that all UTs are in a suburban scenario. 

\begin{figure}[!t]
\centering
\subfloat[]{\includegraphics[width=2.5in]{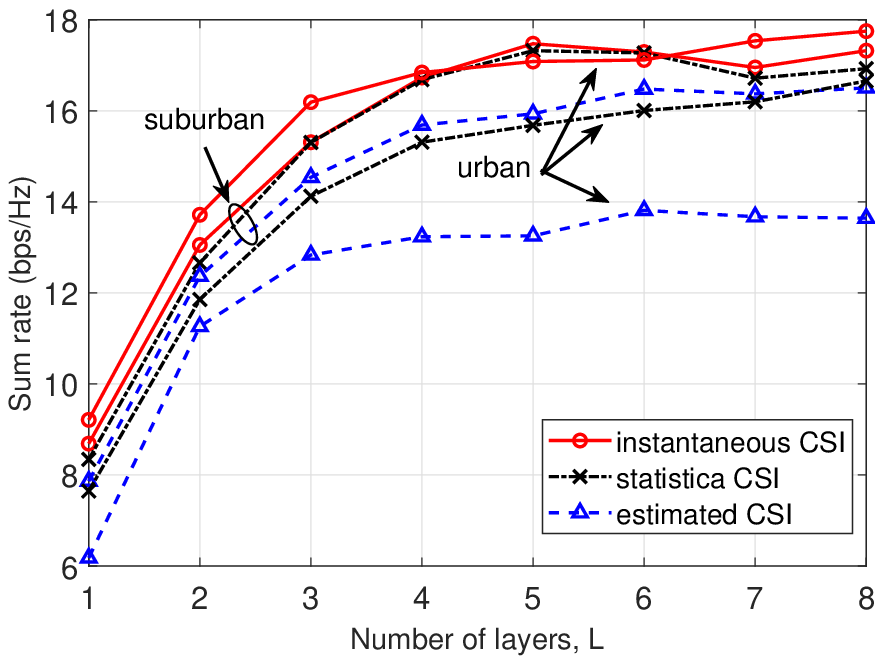}%
\label{fig_2a}}
\hfil
\subfloat[]{\includegraphics[width=2.5in]{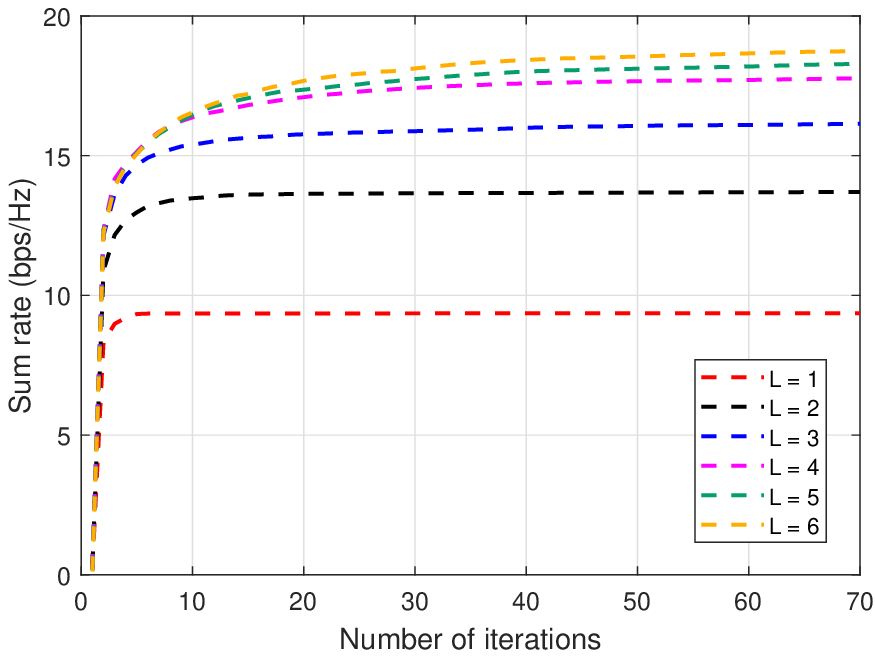}%
\label{fig_2b}}
\caption{(a) Sum rate versus the number of layers $L$ ($P_T=30$ dBW, $N=225$, $M=K=9$). (b) Convergence behavior of the AO algorithm.}
\label{fig_2}
\end{figure}

In Fig. 2(a), we illustrate the sum rate versus the number of SIM layers $L$. Specifically, we compare the performance of the AO algorithm adopting statistical CSI, instantaneous CSI, and estimated CSI in two scenarios: urban and suburban areas, where the Rician factor is generally higher in suburban than in urban. We observed that using statistical CSI shows better performance compared to that relying on estimated instantaneous CSI. And the gain is more pronounced when the Rician factor is smaller. The sum rate increases initially with $L$ and then becomes stable. This is caused by a complicated tradeoff between the increase in the number of metasurface layers and the decrease in inter-layer spacing. The former enhances SIM's beamforming capability, while the latter, reducing the effective energy radiation from one layer to another, degrades the beamforming capability. For a small number of metasurface layers, the former dominates, while with more layers, the latter becomes dominant. Moreover, we also verified the convergence of the AO algorithm for different $L$. As shown in Fig. 2(b), the algorithm can converge as $L$ increases, but the convergence speed decreases.

\begin{figure}[!t]
\centering
\subfloat[]{\includegraphics[width=2.5in]{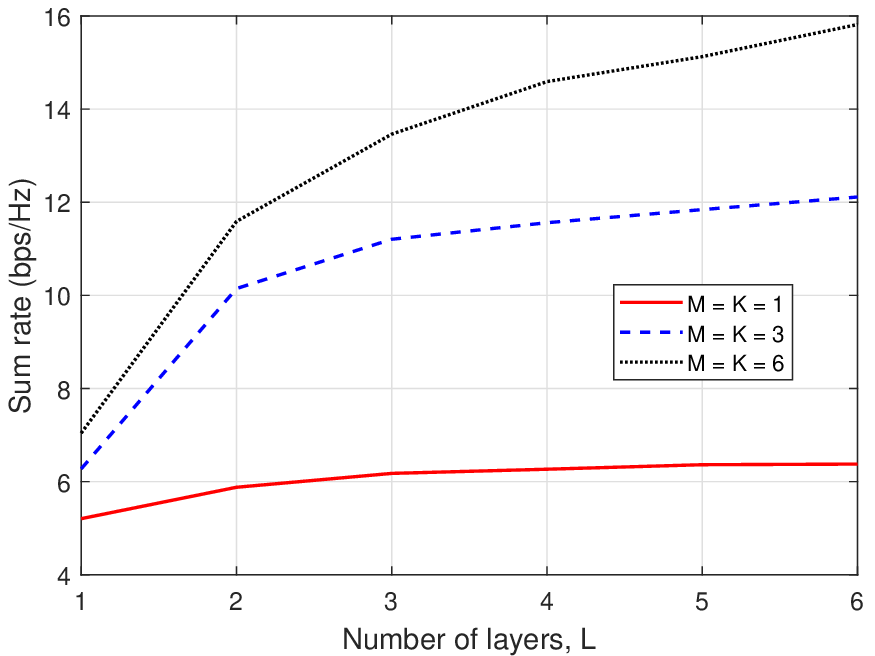}%
\label{fig_3a}}
\hfil
\subfloat[]{\includegraphics[width=2.5in]{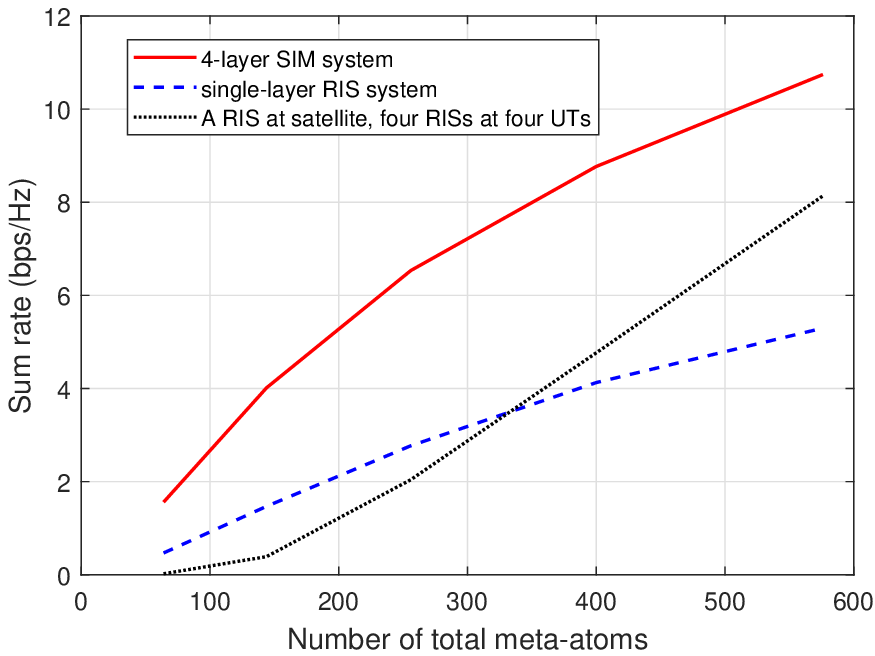}%
\label{fig_3b}}
\caption{(a) Sum rate versus the number of layers $L$ when the number of total meta-atoms is fixed ($P_T=30$ dBW, $LN=1200$). (b) Sum rate versus the number of total meta-atoms ($P_T=30$ dBW, $M=K=4$).}
\label{fig_3}
\end{figure}

In Fig. 3(a), we present the sum rate versus the number of layers $L$ when the number of total meta-atoms is set to 1200. Additionally, we consider three scenarios with different numbers of served UTs $K$. The results indicate that the sum rate increases with the number of UTs $K$ and the number of layers $L$. The former provides multiuser multiplexing gain, while the latter offers enhanced computational capability to mitigate multiuser interference in the wave domain. Furthermore, in Fig. 3(b), we present the sum rate versus the number of total meta-atoms for $M=K=4$. We also consider two systems for comparison: a single-layer RIS system and a dual-layer SIM-like system with the second layer RIS being evenly split on the UT side. All phase shifts are optimized by adapting the proposed AO algorithm. It is observed that the SIM-based system outperforms the benchmarks since multiple layers provide enhanced capability for mitigating the interference. 
\begin{figure}[!t]
\centering
\subfloat[]{\includegraphics[width=2.5in]{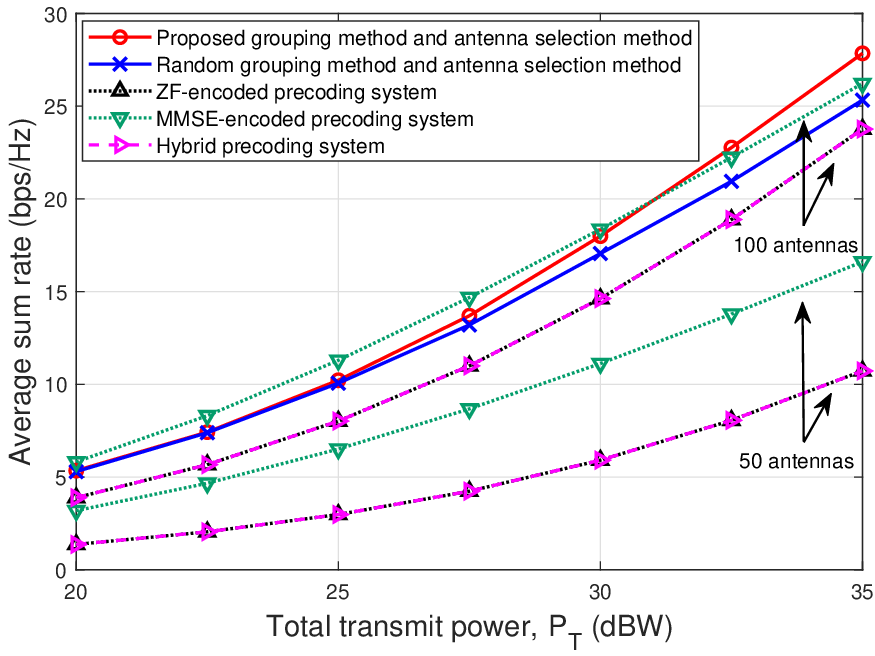}%
\label{fig_4a}}
\hfil
\subfloat[]{\includegraphics[width=2.5in]{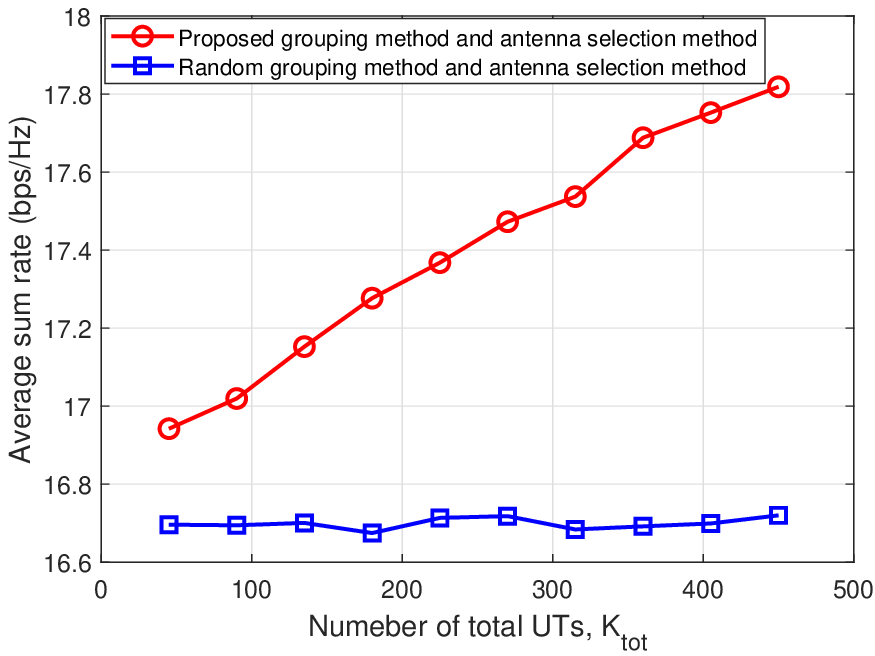}%
\label{fig_4b}}
\caption{(a) Average sum rate versus total transmit power $P_T$ ($N=225$, $L=4$, $K_{\text{tot}}=270$, $M=K=9$). (b) Average sum rate versus the number of total UTs $K_\text{tot}$ ($P_T=30$ dBW, $N=225$, $L=4$, $M=K=9$).}
\label{fig_4}
\end{figure}

 In Fig. 4(a), we illustrate the average sum rate versus the total transmit power $P_T$. Meanwhile, we compare the SIM-based system using the proposed grouping method and antenna selection method with four benchmark schemes: an SIM-based system using random grouping method and random antenna selection method, a digital system based on ZF and MMSE precoding method, and a hybrid precoding scheme \cite{liang2014low}. The latter two adopt the proposed grouping method. To maintain the same CSI conditions, we position antennas at a portion of the last SIM layer's meta-atoms for the digital and hybrid precoding systems. Furthermore, we use the standard water-filling algorithm for power allocation during the deployment of the ZF precoding. The user grouping algorithm provides an average gain of around 4.4\% compared to the random approach. Additionally, SIM-based systems, compared to those relying on digital precoding, achieve similar performance with a significant reduction in the number of RF chains. Furthermore, the performance of our SIM-based system can be further enhanced by increasing the number of low-cost meta-atoms. Moreover, in Fig. 4(b), we illustrate the average sum rate versus the number of total UTs $K_\text{tot}$. Meanwhile, we compare the performance of the proposed user grouping and antenna selection algorithms with the random benchmark schemes. The results show that our approach consistently outperforms the random method. Moreover, we note that the performance gain increases with the growth of $K_\text{tot}$ as it provides a larger selection space for user combinations, resulting in enhanced channel orthogonality within each group.
\section{ Conclusions}
In this letter, we proposed an SIM-based multibeam LEO satellite communication system. Compared to traditional multi-beam satellite systems based on digital precoding, SIM performs downlink precoding in the wave domain, resulting in lower processing latency and computational burden. Furthermore, by employing the statistical CSI, we formulated an optimization problem to maximize the ergodic sum rate by designing phase shifts of the SIM and transmit power allocation. A customized AO algorithm was proposed to effectively solve this joint optimization problem. Additionally, we proposed a user grouping method based on the CoC minimization principle and designed an antenna selection algorithm by minimizing the total leaked energy. The simulation results demonstrated the effectiveness of the proposed SIM-based design and the customized AO algorithm compared to various benchmarks. In a nutshell, we explore a new wave-based beamforming paradigm in the design of multibeam satellite systems to carry out downlink precoding with ultra-fast computational speed.

\bibliographystyle{IEEEtran}
\bibliography{reference}

% Generated by IEEEtran.bst, version: 1.14 (2015/08/26)
\begin{thebibliography}{10}
\providecommand{\url}[1]{#1}
\csname url@samestyle\endcsname
\providecommand{\newblock}{\relax}
\providecommand{\bibinfo}[2]{#2}
\providecommand{\BIBentrySTDinterwordspacing}{\spaceskip=0pt\relax}
\providecommand{\BIBentryALTinterwordstretchfactor}{4}
\providecommand{\BIBentryALTinterwordspacing}{\spaceskip=\fontdimen2\font plus
\BIBentryALTinterwordstretchfactor\fontdimen3\font minus \fontdimen4\font\relax}
\providecommand{\BIBforeignlanguage}[2]{{%
\expandafter\ifx\csname l@#1\endcsname\relax
\typeout{** WARNING: IEEEtran.bst: No hyphenation pattern has been}%
\typeout{** loaded for the language `#1'. Using the pattern for}%
\typeout{** the default language instead.}%
\else
\language=\csname l@#1\endcsname
\fi
#2}}
\providecommand{\BIBdecl}{\relax}
\BIBdecl

\bibitem{arXiv_2023_An_Toward}
\BIBentryALTinterwordspacing
J.~An, C.~Yuen, L.~Dai, M.~Di~Renzo, M.~Debbah, and L.~Hanzo, ``Toward beamfocusing-aided near-field communications: Research advances, potential, and challenges,'' 2023. [Online]. Available: \url{https://arxiv.org/pdf/2309.09242.pdf}
\BIBentrySTDinterwordspacing

\bibitem{kodheli2020satellite}
O.~Kodheli, E.~Lagunas, N.~Maturo, S.~K. Sharma, B.~Shankar, J.~F.~M. Montoya, J.~C.~M. Duncan, D.~Spano, S.~Chatzinotas, S.~Kisseleff \emph{et~al.}, ``{Satellite communications in the new space era: A survey and future challenges},'' \emph{IEEE Commun. Surv. Tuts.}, vol.~23, no.~1, pp. 70--109, Oct. 2020.

\bibitem{schwarz2019mimo}
R.~T. Schwarz, T.~Delamotte, K.-U. Storek, and A.~Knopp, ``{MIMO applications for multibeam satellites},'' \emph{IEEE Trans. Broadcast.}, vol.~65, no.~4, pp. 664--681, Dec. 2019.

\bibitem{li2021downlink}
K.-X. Li, L.~You, J.~Wang, X.~Gao, C.~G. Tsinos, S.~Chatzinotas, and B.~Ottersten, ``{Downlink transmit design for massive MIMO LEO satellite communications},'' \emph{IEEE Trans. Commun.}, vol.~70, no.~2, pp. 1014--1028, Nov. 2021.

\bibitem{TGCN_2022_An_Joint}
J.~An, C.~Xu, L.~Wang, Y.~Liu, L.~Gan, and L.~Hanzo, ``Joint training of the superimposed direct and reflected links in reconfigurable intelligent surface assisted multiuser communications,'' \emph{IEEE Trans. Green Commun. Netw.}, vol.~6, no.~2, pp. 739--754, Jun. 2022.

\bibitem{chen2022robust}
Y.~Chen, Y.~Wang, and L.~Jiao, ``{Robust transmission for reconfigurable intelligent surface aided millimeter wave vehicular communications with statistical CSI},'' \emph{IEEE Trans. Wireless. Commun.}, vol.~21, no.~2, pp. 928--944, Feb. 2022.

\bibitem{AnChannel}
J.~An, C.~Xu, L.~Gan, and L.~Hanzo, ``Low-complexity channel estimation and passive beamforming for {RIS}-assisted {MIMO} systems relying on discrete phase shifts,'' \emph{IEEE Trans. Commun.}, vol.~70, no.~2, pp. 1245--1260, Feb. 2022.

\bibitem{AnCodebook}
J.~An, C.~Xu, Q.~Wu, D.~W.~K. Ng, M.~D. Renzo, C.~Yuen, and L.~Hanzo, ``Codebook-based solutions for reconfigurable intelligent surfaces and their open challenges,'' \emph{IEEE Wireless. Commun.}, pp. 1--8, Jun. 2022.

\bibitem{AnAntenna}
C.~Xu, J.~An, T.~Bai, S.~Sugiura, R.~G. Maunder, L.-L. Yang, M.~Di~Renzo, and L.~Hanzo, ``Antenna selection for reconfigurable intelligent surfaces: A transceiver-agnostic passive beamforming configuration,'' \emph{IEEE Trans. Wireless. Commun.}, pp. 1--1, Mar. 2023.

\bibitem{an2023stacked}
J.~An, M.~Di~Renzo, M.~Debbah, and C.~Yuen, ``{Stacked intelligent metasurfaces for multiuser beamforming in the wave domain},'' in \emph{Proc. IEEE Int. Conf. Commun. (ICC)}, Rome, Italy, May 2023, pp. 1--6.

\bibitem{AnSIM}
J.~An, C.~Xu, D.~W.~K. Ng, G.~C. Alexandropoulos, C.~Huang, C.~Yuen, and L.~Hanzo, ``Stacked intelligent metasurfaces for efficient holographic {MIMO} communications in 6{G},'' \emph{IEEE J. Sel. Areas Commun.}, vol.~41, no.~8, pp. 2380--2396, Aug. 2023.

\bibitem{an2023stackeds}
J.~An, C.~Yuen, C.~Xu, H.~Li, D.~W.~K. Ng, M.~Di~Renzo, M.~Debbah, and L.~Hanzo, ``{Stacked intelligent metasurface-aided MIMO transceiver design},'' \emph{arXiv preprint arXiv:2311.09814}, 2023.

\bibitem{lin2018all}
X.~Lin, Y.~Rivenson, N.~T. Yardimci, M.~Veli, Y.~Luo, M.~Jarrahi, and A.~Ozcan, ``{All-optical machine learning using diffractive deep neural networks},'' \emph{Science}, vol. 361, no. 6406, pp. 1004--1008, Jul. 2018.

\bibitem{arXiv_2024_An_Two}
\BIBentryALTinterwordspacing
J.~An, C.~Yuen, Y.~L. Guan, M.~Di~Renzo, M.~Debbah, H.~V. Poor, and L.~Hanzo, ``Two-dimensional direction-of-arrival estimation using stacked intelligent metasurfaces,'' 2024. [Online]. Available: \url{https://arxiv.org/pdf/2402.08224.pdf}
\BIBentrySTDinterwordspacing

\bibitem{jindal2005sum}
N.~Jindal, W.~Rhee, S.~Vishwanath, S.~A. Jafar, and A.~Goldsmith, ``{Sum power iterative water-filling for multi-antenna Gaussian broadcast channels},'' \emph{IEEE Trans. Inf. Theory}, vol.~51, no.~4, pp. 1570--1580, Apr. 2005.

\bibitem{hu2017user}
M.~Hu, Y.~Chang, T.~Zeng, X.~Yang, and A.~Men, ``{User grouping with adaptive group number for massive MIMO downlink systems},'' in \emph{Proc. 20th Int. Symp. Wireless Pers. Multimedia Commun. (WPMC)}, Bali, Indonesia, Dec. 2017, pp. 459--464.

\bibitem{kuhn1955hungarian}
H.~W. Kuhn, ``{The Hungarian method for the assignment problem},'' \emph{Nav. Res. Logist.}, vol.~2, no. 1-2, pp. 83--97, Mar. 1955.

\bibitem{3GPPTR}
``{\textit{Study on New Radio (NR) to Support Non-Terrestrial Networks (Release 15)}},'' 3GPP, Sophia Antipolis, Valbonne, France, {document TR 38.811, V15.4.0}, Sep. 2020.

\bibitem{CL_2023_An_A}
J.~An, C.~Yuen, C.~Huang, M.~Debbah, H.~Vincent~Poor, and L.~Hanzo, ``A tutorial on holographic mimo communications—part {I}: Channel modeling and channel estimation,'' \emph{IEEE Commun. Lett.}, vol.~27, no.~7, pp. 1664--1668, Jul. 2023.

\bibitem{liang2014low}
L.~Liang, W.~Xu, and X.~Dong, ``{Low-complexity hybrid precoding in massive multiuser MIMO systems},'' \emph{IEEE Wireless. Commun. Lett.}, vol.~3, no.~6, pp. 653--656, Oct. 2014.

\end{thebibliography}
\vfill
\end{document}